# Thermal Contrast in Nanoscale Infrared Spectroscopy (AFM-IR): Low Frequency Limit

Anna N. Morozovska[1,2], Eugene A. Eliseev,[3] N. Borodinov,[4] O. Ovchinnikova,[5] Nicholas V. Morozovsky[1] and Sergei V. Kalinin[5, *]

[1]*Institute of Physics, National Academy of Sciences of Ukraine,*
*46, Prospekt Nauky, 03028 Kyiv, Ukraine,*

[2]*Bogolyubov Institute for Theoretical Physics, National Academy of Sciences of Ukraine,*
*14-b Metrolohichna str. 03680Kyiv, Ukraine*

[3]*Institute for Problems of Materials Science, National Academy of Sciences of Ukraine,*
*3, Krjijanovskogo, 03142 Kyiv, Ukraine*

[4]*Department of Materials Science and Engineering, Clemson University,Clemson,*
*South Carolina 29634, USA,*

[5]*Center for Nanophase Materials Science, Oak Ridge National Laboratory, Oak Ridge,*
*Tennessee 37831, USA*

The contrast formation mechanism in Nanoscale Infrared Spectroscopy (Nano-IR or AFM-IR) is analyzed for the boundary between two layers with different light absorption, thermal and elastic parameters. Analytical results derived in the decoupling approximation for low frequency limit show that the response amplitude is linearly proportional to the intensity of the illuminating light and thermal expansion coefficient. The spatial resolution between two dissimilar materials is linearly proportional to the sum of inverse light adsorption coefficients and to the effective thermal transfer length. The difference of displacements height across the T-shape boundary ("thermo-elastic step") is proportional to the difference of the adsorption coefficients and inversely proportional to the heat transfer coefficient. The step height becomes thickness-independent for thick films and proportional to $h^2$ in a very thin film.

---

[*] Corresponding author. E-mail: sergei2@ornl.gov (S.V.K.)

Infrared spectroscopy is a well-explored analytical technique that finds numerous application is chemistry,[1, 2] on-line monitoring sensors[3] and industrial set-ups.[4] It is based on a wavelength-specific absorption of light due to the molecular vibrations. These spectra contain[5-7] information regarding chemical composition, molecular orientation, crystallinity and defects of the materials structure. The ability to analyze it locally gives new insights on the distribution of the absorbers within the sample. The conventional IR microscopy allows for the spatial resolution in the range of microns[8], which is limited by the optical diffraction. However, by using the atomic force microscopy it is possible to significantly scale down the size of the region being probed. In this AFM-IR set-up the infrared light is absorbed by the sample and the sharp tip detects the mechanical displacement the originating from the thermal expansion. As a result, the local properties of the various objects as well as the mapping capabilities are available to researchers working on polymers,[9, 10] biological samples[11, 12] and semiconductors.[13, 14]

There has been a significant progress over the course of last five years towards the improvement of the AFM-IR as a modern technique of analysis. However, the contrast formation mechanism in this technique remains relatively unknown. The case of a spherical sample surrounded by the isotropic homogenous media was examined,[15] with the special attention has been paid to the dynamics of the cantilever as it comes in contact with the periodic photothermally induced sample expansion. However, further progress in the nanoscale infrared analysis required rigorous investigation of the fundamental relationships between physical parameters of the sample and the quality of the spectral acquisition and mapping.

This paper focuses on the on the analytical treatment of the photothermal expansion of the contacting region of two materials with the varying infrared absorption coefficients located on a rigid substrate. The contrast between the regions and AFM-IR scan is being considered in details. The width of the thermoelastic response transient region and the step height are being calculated as functions of the inverse light adsorption coefficients, effective thermal transfer length and the thickness of the film. This analysis follows the previous development of linear resolution theory for piezoresponse force microscopy[16-19] and electrochemical strain microscopy.[20]

Specifically, our model considers the film of thickness $h$ on the infinitely thick rigid substrate. This film is comprised of region A and B separated by initially vertical boundary, deposited on substrate C [**Fig. 1(a)**]. The elastic properties, heat conductions, and light adsorption lengths of the regions A, B and C are known. The surface is illuminated by the periodically modulated **IR**-illumination, while the modulation frequency ($\omega_0$=1 kHz – 1 MHz) is well below IR frequencies. Light adsorbs according to the Beer's law and generates heat hence causing thermal expansion. The latter causes a bulk strain leading to the surface deformation. We aim to calculate the mechanical displacement profile across A-B boundary caused by the light illumination that can be measured experimentally via AFM tip [**Fig. 1(b)**]. Here, we do not consider other light-induced mechanical responses (photostriction, etc.), and do not consider the cantilever dynamics (that can be readily analyzed via standard transfer function theory[21]).

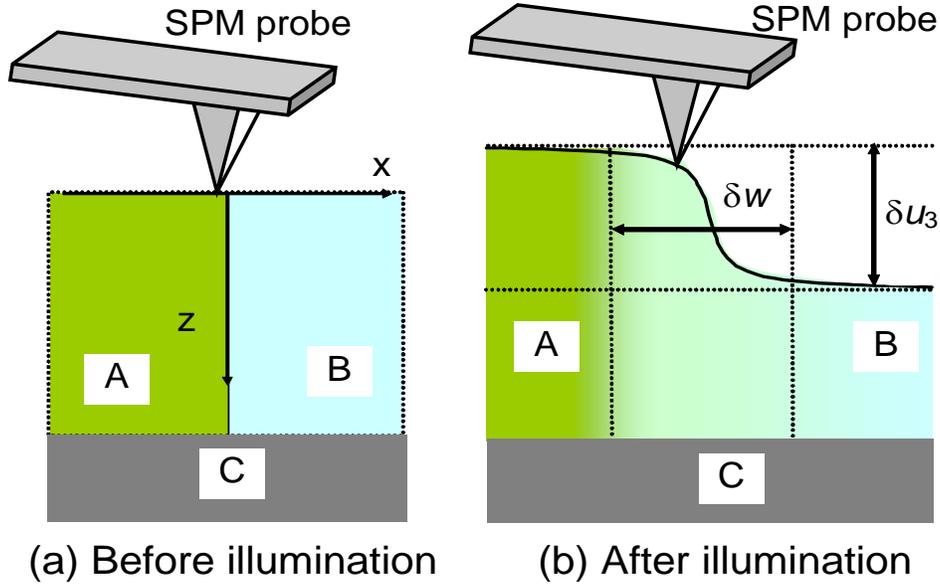

(a) Before illumination    (b) After illumination

**FIG.1 (a)** Initial state before the light illumination. **(b)** After illumination.

**I. Theoretical formalism. A. Heat sub-problem.** Temperature distributions in the multi-region system is described by a coupled system of linear thermal conductivity equations, at that each layer is characterized by its own equation for the temperature variation $\vartheta_m(x,z,t)$ inside each region:

$$c_m \frac{\partial}{\partial t} \vartheta_m = k_m^T \left( \frac{\partial^2}{\partial x^2} + \frac{\partial^2}{\partial z^2} \right) \vartheta_m + q_m(x,z,t), \tag{1}$$

where $m = A, B, C$ corresponds to the regions A ($x \leq 0$, $0 \leq z \leq h$ at $t$=0), B ($x \geq 0$, $0 \leq z \leq h$ at $t$=0) and C ($-\infty < x < \infty$, $z > h$). Note that the physical boundary $S_{AB}$ between the regions A and B can shift with time as well as the top surface $z = 0$ bends due to the thermo-elastic effects, and numerical simulations account for all these changes in a self-consistent way. The thermal diffusivities $\kappa_m = k_m^T/c_m$, where $c_m$ is the heat capacity and $k_m^T$ is the thermal conductivity of the region "$m$". Assuming that all thermal sources $q_m(x,z,t)$ are proportional to the light intensity distribution given by the Beer's law, $q_m(x,z,t) = \gamma_m I_0(x,t)\exp(-\alpha_m z)$, where $z \geq 0$, light absorption coefficients $\alpha_m$ and prefactors $\gamma_m I_0$ are different in the regions A, B and C. Here we consider $I_0(x,t) = I_0(t)\exp(-\lambda|x|)$, where the decay constant $0 \leq \lambda \ll \alpha$, i.e. change slowly.

The thermal boundary conditions to Eqs. (1) on the top surface have the form of heat balance equations with different heat exchange coefficients $g_m$ between the top surfaces of the regions A and B and environment,[22] namely the normal heat fluxes $j_3^m = -k_m^T(\partial\vartheta_m/\partial\vec{n})\big|_{S_m} = -g_m \vartheta_m(x,z,t)$ ($m$=A, B). The thermal boundary conditions at the physical boundaries AB, AC and BC are the continuity of heat fluxes and the equality of the layers' temperatures, e.g. $k_A^T \dfrac{\partial\vartheta_A}{\partial\vec{n}} - k_B^T \dfrac{\partial\vartheta_B}{\partial\vec{n}}\bigg|_{S_{AB}} = 0$, $\vartheta_A - \vartheta_B\big|_{S_{AB}} = 0$.

**B. Thermoelastic sub-problem.** Generalized linear Hooke's law relating elastic stresses and strains in a thermo-elastic problem for each region A, B, C has the form $\sigma_{pq} = c_{pqij}^m(u_{ij} - \beta_{ij}^m \vartheta_m)$, where $m$=A, B, C without summation; $u_{ij}$ is the strain tensor, $c_{pqij}^m$ and $s_{ijkl}^m$ are elastic stiffness and compliances tensors, $\beta_{ij}^m$ is the thermal expansion tensor. Hereinafter $x = x_1$, $y = x_2$ and $z = x_3$. We note that the typical contact area in SPM experiment is well below micron-scale. The corresponding intrinsic resonance frequencies of material are thus in the GHz range, well above the practically important limits (both in terms of ion dynamic, and SPM-based detection of localized mechanical vibrations). Hence, we can the equation of mechanical equilibrium in the quasi-static case. This mechanical equilibrium equation, $\partial\sigma_{ij}/\partial x_j = 0$, along with Hooke's law leads to the Lame-type equations for mechanical displacement vector $U_k$ in the regions A, B, C:

$$c_{ijkl}^m \left( \dfrac{\partial^2 U_k}{\partial x_j \partial x_l} - \beta_{kl}^m \dfrac{\partial \vartheta_m}{\partial x_j} \right) = 0. \qquad (2)$$

Elastic boundary conditions to Eqs.(2) are the absence of the normal stresses mechanically free surfaces of media "A" and "B" $\left(\sigma_{pq}n_q\right)\big|_{S_m} \equiv c_{pqij}^m n_q \left(u_{ij}^m - \beta_{ij}^m \vartheta_m\right)\big|_{S_m} = 0$ ($m = A, B$). At the interface between "A" and "B" conditions of continuity of normal stresses $\left(\sigma_{pq}n_q\right)\big|_{S_{AB}-0} = \left(\sigma_{pq}n_q\right)\big|_{S_{AB}+0}$ and mechanical displacement $U_i\big|_{S_{AB}-0} = U_i\big|_{S_{AB}+0}$ should be satisfied. At the boundary with matched substrate elastic displacements should be continuous $U_i^A\big|_{S_{AC}} = U_i^B\big|_{S_{BC}}$. The displacement is zero on the boundary with the rigid substrate $U_i^A\big|_{S_{AC}} = U_i^B\big|_{S_{BC}} = 0$. The boundary problem (1)-(2) has been solved numerically, and results will be presented below.

**II. Analytical solution in decoupling approximation.** To develop analytical solution, we employ the decoupled approximation.[18, 23] In this case, it is assumed that the solution of the thermal problem yields thermal field inside the material. The temperature change generates stress field, from which the displacement can be calculated. However, the mechanical displacement does not affect the solution of the thermal problem. This approach also allows using approximations for the solution of the individual problems, for example changing the symmetries of different parameter tensors and introducing simplifying approximations on material properties as shown previously for piezoresponse force microscopy[16-19, 21, 24-26] and electrochemical strain microscopy.[20]

To derive tractable analytical solution for thermal problem, we assume that all thermal properties of the semi-infinite ($h \to \infty$) media A and B are the same ($c_A \approx c_B \approx c_F$, $k_A^T \approx k_B^T \approx k_F^T$, $g_A \approx g_B \approx g_F$), except for the light adsorption coefficients ($\alpha_A \neq \alpha_B$) and factors ($\gamma_A \neq \gamma_B$), and substrate that does not contain heat sources. For the case the thermal boundary condition at the boundary $S_{AB}$ (3c) does not matter, and the T-type geometry of the thermal problem reduces to the simple two-layer problem with inhomogeneous thermal source. For the case the Fourier image of the temperature field becomes essentially simpler (see electronic supplementary materials), namely:

$$\tilde{\vartheta}(k, z, \omega) = A(k, \omega)\exp(-kz) + \tilde{\vartheta}_P(k, z, \omega), \tag{3a}$$

$$A(k, \omega) = \frac{1}{g_F + k_F^T |k|}\left(k_F^T \frac{\partial \tilde{\vartheta}_P(k, 0, \omega)}{\partial z} - g_F \tilde{\vartheta}_P(k, 0, \omega)\right), \tag{3b}$$

Explicit form of $\tilde{\vartheta}_P(k, z, \omega)$ is

$$\tilde{\vartheta}_P(k, z, \omega) = -\frac{\tilde{I}_0(\omega)}{\sqrt{2\pi} k_F^T} \left( \frac{\gamma_A \exp(-\alpha_A z)}{(\lambda + ik)(\alpha_A^2 - k^2 - i\omega\kappa_F)} + \frac{\gamma_B \exp(-\alpha_B z)}{(\lambda - ik)(\alpha_B^2 - k^2 - i\omega\kappa_F)} \right). \quad (3c)$$

The displacement can be found in decoupled approximation assuming that all elastic properties of the media A and B are the same ($c^A_{pqij} \approx c^B_{pqij}$), and both materials are placed on elastically matched substrate (i.e. the latter assumption is a very good approximation for the case when the materials A and B are in fact one material doped by different photoactive impurities (including the particular cases of one doped and other pristine materials). In these approximations the elastic boundary between the media A and B virtually disappears, only boundary conditions at $x=0$ and $x=h$ matter. For the isotropic thermal expansion tensor, $\beta_{11} = \beta_{22} = \beta_{33} = \beta$, we derived [20] the vertical surface displacement induced by the redistribution of temperature $\vartheta(x_1, x_2, x_3, t)$:

$$\tilde{u}_3(k, h, \omega) = \frac{(1+\nu)\beta \tilde{I}_0(\omega)}{\pi\sqrt{2\pi} k_F^T} \left( \frac{\gamma_A}{(\lambda + ik)} F(k, h, \omega, \alpha_A) + \frac{\gamma_B}{(\lambda - ik)} F(k, h, \omega, \alpha_B) \right), \quad (4a)$$

$$F(k, h, \omega, \alpha) = \frac{1 - \exp[-(\alpha + |k|)h]}{(\alpha + |k|)(\alpha^2 - k^2 - i\omega\kappa_F)} - \frac{g_F + k_F^T \alpha}{g_F + k_F^T |k|} \frac{1 - \exp(-2|k|h)}{2|k|(\alpha^2 - k^2 - i\omega\kappa_F)}. \quad (4b)$$

Here $\nu$ is a Poisson ratio. Elementary estimates show that the low frequency limit ($\omega \to 0$) can be a reasonable approximation when the absorption coefficients are enough high in the actual frequency range, i.e. $\omega \ll \alpha_m^2/\kappa_F$. Being further interested in the case $\omega = 0$ for which the seeming pole exists in Eq.(4b), we simplify the expression (4a) as

$$\tilde{u}_3(k, h) \approx \frac{(1+\nu)\beta I_0}{\pi\sqrt{2\pi} k_F^T} \frac{1}{(g_F/k_F^T) + |k|} \left( \frac{\gamma_A(1 - \exp(-2\alpha_A h))}{2\alpha_A(\lambda + ik)(\alpha_A + |k|)} + \frac{\gamma_B(1 - \exp(-2\alpha_B h))}{2\alpha_B(\lambda - ik)(\alpha_B + |k|)} \right) \quad (5)$$

The expression (5) can be analyzed and simplified in the case of zero flux on the top surface (i.e. assuming that $g_F \to 0$, corresponding to the absence of heat contact, zero heat flux at z=0) and in the opposite case of zero temperature (i.e. at $k_F^T \to 0$). As one can see the surface displacement (5) either diverges for $g_F \to 0$ due to the pole $\sim 1/|k|$ or becomes very small at $k_F^T \to 0$. Hence the realistic situation when both parameters $g_F$ and $k_F^T$ are nonzero should be considered below.

Inverse Fourier transformation of Eq.(5) gives the displacement x-profile in the form:

$$u_3(x,h) = \frac{(1+\nu)\beta I_0}{\pi\sqrt{2\pi}k_F^T}\left[u_f\left(x,h,\gamma_A,\alpha_A,\frac{g_F}{k_F^T},\lambda\right) - u_f\left(x,h,\gamma_B,\alpha_B,\frac{g_F}{k_F^T},-\lambda\right)\right], \quad (6)$$

where the function $u_f$ is introduced:

$$u_f(x,h,\gamma,\alpha,\mu,\lambda) = \frac{\gamma}{2\alpha}[1-\exp(-2\alpha h)]\frac{f(x,\alpha,\lambda) - f(x,\mu,\lambda)}{\alpha - \mu} \quad (7)$$

the main terms of the function $f$ are

$$f(x,\alpha,\lambda) \sim \frac{1}{\sqrt{2\pi}\alpha}\left[\pi(\text{sign}(x)-1)\exp(\lambda x) - \pi\cos(\alpha x)(\text{sign}(x)-\tanh(\alpha x)) - \frac{2\alpha x}{(\alpha x)^2 + (2/(\pi-2))}\right]$$

(8)

The profile (6) saturation obeys long-range Lorentz law [the last term], it is not exponential as anticipated.

The function step height can be estimated as $f(-\infty,\alpha,0) - f(\infty,\alpha,0) \approx -\sqrt{\pi/2}/\alpha$ at $\lambda \to 0$. Using the estimate in Eq.(8) and assuming that $\gamma_A \cong \alpha_A$ and $\gamma_B \cong \alpha_B$ in the case of negligible reflection the in accordance with the energy conservation law, the difference of displacements height ("step") across the boundary AB, $\delta u_3 = u_3(x \to -\infty) - u_3(x \to +\infty)$ [see scheme in **Fig.1(b)**], can be estimated as

$$\frac{(1+\nu)\beta I_0}{4\pi g_F}\left[\frac{1-e^{-2\alpha_A h}}{\alpha_A} - \frac{1-e^{-2\alpha_B h}}{\alpha_B}\right] \approx \frac{(1+\nu)\beta I_0}{4\pi g_F}\begin{cases}\frac{h^2}{2}(\alpha_A - \alpha_B), & h \ll \min[\alpha_A^{-1},\alpha_B^{-1}] \\ \frac{\alpha_B - \alpha_A}{\alpha_A \alpha_B}, & h \gg \max[\alpha_A^{-1},\alpha_B^{-1}]\end{cases} \quad (9)$$

Where $\lambda \to 0$ when deriving Eq.(9). As one can see the step is proportional to the difference of the adsorption coefficients $(\alpha_A - \alpha_B)$ and inversely proportional to the heat transfer coefficient $g_F$; it is thickness-independent for thick films and proportional to $h^2$ in very thin films. However Eq.(5) can be valid for a very thin film only qualitatively because the thermal field becomes essentially different from the semi-infinite mediums A, B, and rigorous numerical calculations should be performed for the case.

Using Eqs.(8) the width $\delta w$ of the thermoelastic response transient region located near the initial boundary AB ($x=0$) [see scheme in **Fig.1(b)**] can be calculated. The width at 50% of maximum amplitude can be estimated as

$$\delta w \approx 5\left(\frac{1}{\alpha_A} + \frac{1}{\alpha_B} + \frac{2k_F^T}{g_F}\right) \qquad (10)$$

Notably, the width of the transient region is linearly proportional to the sum of inverse adsorption coefficients and the to "effective thermal transfer" length $k_F^T/g_F$.

Profiles of vertical surface displacement $u_3(x)$ across the boundary AB calculated in a thick film ($h >> \max[\alpha_A^{-1}, \alpha_B^{-1}]$) with different ratios of light adsorption coefficients $\alpha_B/\alpha_A$ and effective thermal transfer parameter $(\alpha_A k_F^T)/g_F$ are shown in **Figs. 2.** Notably that the step-like changes of the profile are pronounced in the case when the ratio $\alpha_B/\alpha_A$ strongly differs from unity [**Fig.2(a)**].

Small ratio $(\alpha_A k_F^T)/g_F$ corresponds to the enough sharp step of $u_3(x)$ at fixed ratio $\alpha_B/\alpha_A$, but the total amplitude of the surface displacement is rather small [**Fig.2(b)**]. Big ratio $(\alpha_A k_F^T)/g_F$ corresponds to the rather smeared profiles with wide transient regions, but total amplitude of the surface displacement is rather high [**Fig.2(c)**].

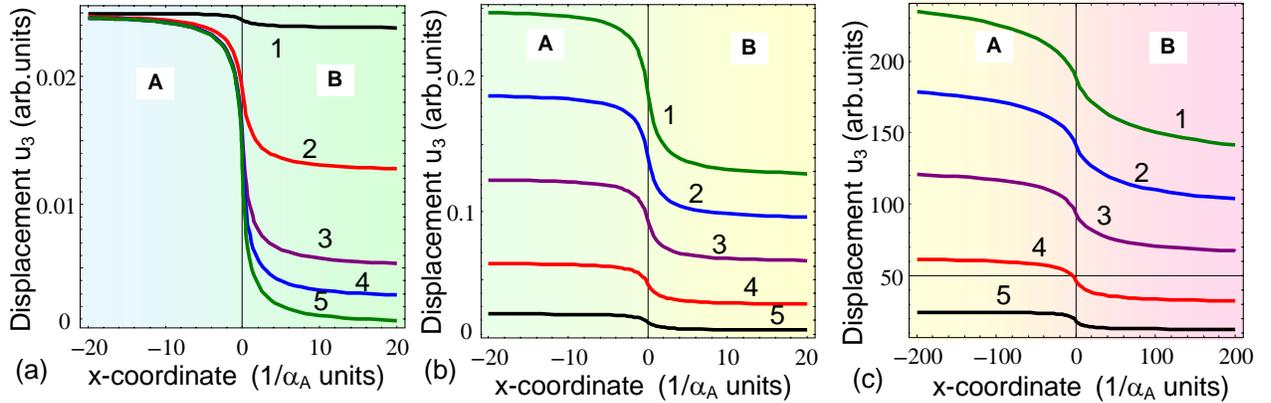

**FIG. 2.** Profiles of vertical surface displacement across the boundary AB calculated at different light adsorption coefficients ratio $\alpha_B/\alpha_A$= 1.05, 2, 5, 10, 50 and fixed $(\alpha_A k_F^T)/g_F$ =0.01 (curves 1-5 in the plot (**a**)); different ratio $(\alpha_A k_F^T)/g_F$ =0.1, 0.075, 0.05, 0.025, 0.01 (curves 1-5 in the plot (**b**)), $(\alpha_A k_F^T)/g_F$ =100, 75, 50, 25, 10 (curves 1-5 in the plot (**c**)) at fixed $\alpha_B/\alpha_A$=2. Parameter $\lambda/\alpha_A$= $10^{-4}$.

The physical picture [shown in **Fig.2**] approves the conclusions made from approximate expressions (6)-(8), namely:

a)   thermoelastic response amplitude is linearly proportional to the intensity of the illuminating light and thermal expansion coefficient in decoupling approximation.

b)   the width of the thermoelastic response transient region is linearly proportional to the sum of inverse adsorption coefficients and the to "effective thermal transfer" length $k_F^T/g_F$ ;

c)   the difference of displacements height across the boundary AB ("thermo-elastic step") is proportional to the difference of the adsorption coefficients $(\alpha_A - \alpha_B)$ and inversely proportional to the heat transfer coefficient $g_F$ ;

d)   step nontrivially depends on the film thickness h. It is h-independent for thick films and proportional to $h^2$ for very thin films.

Note that results of this section can be valid for a very thin film only qualitatively because the thermal field becomes essentially different from the semi-infinite mediums A, B. Results of rigorous numerical calculations performed for thin films are presented in next section.

**III. Results of numerical modeling for the static response.** To go beyond the linear decupling approximation we solved numerically the coupled problem that statement is given by Eqs.(1)-(6). Results of the finite elements modeling (FEM) are presented below for two cases. We consider strongly or slightly doped or pristine PMMA material as the regions A and B, which differs only in the light absorption coefficients [see the first column in **Table I**]. We also regard that $\gamma_m \cong \alpha_m$ in accordance with the energy conservation law in the case of negligible light reflection from the surface z=0.

**Table I.** Parameters used in the simulations

| Parameter   (dimensionality) | Regions | | |
|---|---|---|---|
| | **Region A (PMMA)** | **Region B (PMMA)** | **Substrate C (Si)** |
| density (kg/m$^3$) | 1188  [a] | 1188  [a] | 2329  [a] |
| heat capacity $c_m$ (J/(kg K)) | 1460  [b] | 1460  [b] | 718  [c] |
| thermal conductivity $k_m^T$ W/(m·K) | 0.2  [a] | 0.2  [a] | 149  [a] |
| heat exchange coefficient $g_m$ W/(K m$^2$) | $10^3 – 10^5$ | $10^3 – 10^5$ | NA |
| light absorption coefficient $\alpha_m$ (m$^{-1}$) at 3.5 μm wavelength of IR radiation | 3.5×10$^4$  [d] | (3.5 – 350) ×10$^6$ depending on doping | (20 - 3×10$^4$)  [e] depending on doping |
| thermal expansion $\beta_{ij}^m$ (10$^{-6}$ K$^{-1}$) | 100  [f] | 100  [f] | 2.6  [c] |
| Young modulus (GPa) | 3.33 [a] | 3.33 [a] | 188 [a] |
| Poisson ratio (dimensionless) | –0.03 [a] | –0.03 [a] | 0.28 [a] |
| Reduced light intensity $I_0$ ( W/mm$^2$) | 1 | 1 | N/A |

[a] Springer Handbook, page 828
[b] http://www.mit.edu/~6.777/matprops/pmma.htm
[c] http://periodictable.com/Elements/014/data.html
[d] Paul W. Kruse, Laurence D. McGlauchlin, and Richmond B. McQuistan. "Elements of infrared technology: Generation, transmission and detection." *New York: Wiley, 1962* (1962).
[e] W. Spitzer, and H. Y. Fan. "Infrared absorption in n-type silicon." *Physical Review* 108, no. 2 (1957): 268.;  S. C. Shen, C. J. Fang, M. Cardona, and L. Genzel. "Far-infrared absorption of pure and hydrogenated a-Ge and a-Si." Physical Review B 22, no. 6: 2913 (1980).
[f] http://www.tangram.co.uk/TI-Polymer-PMMA.html

We chose PMMA because of its perfect compatibility with many other advanced materials to forming polymer-polymer, polymer-oxide, polymer-semiconductor and nano-composite thin-film diphasic structures on Si-substrates,[27-31] which are in particular promising for nanoscale memory charge switching devices used conducting atomic force microscopy-Kelvin probe microscopy writing-reading technique [27].

The temperature and strain fields inside the 1-µm layers of heavily doped with a photoactive impurity (region B on the right) and slightly doped or pure PMMA (region B on the left) deposited on a transparent silicon substrate are shown in **Fig. 3(a)** and **3(b),** respectively. Since the light absorption coefficients of doped and pure PMMA differ by a factor of 100, the entire heating (by 3.5 K) occurs in the highly doped PMMA region that creates a "hump-like" surface deformation of the order of 0.035% leading to an inhomogeneous rising of the surface of the parts of the layer. The maximal height of the hump is on the order of 200 pm far from the boundary AB.

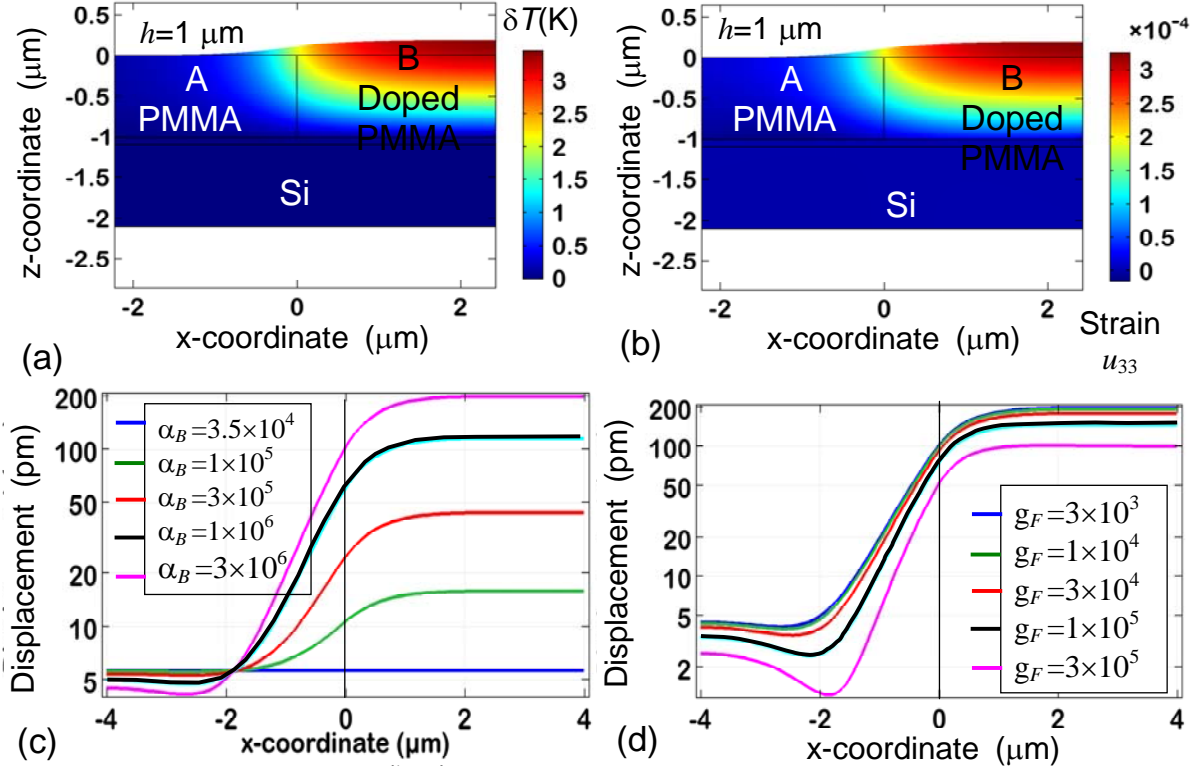

**FIG. 3.** Distributions of temperature variation $\vartheta(x,z)$ **(a)** and strain component $u_{33}(x,z)$ **(b)** in the cross-section of 1 μm thick PMMA films on Si substrate calculated for the light intensity $I_0 = 1$ W/mm², absorption coefficients $\alpha_A = 3.5 \times 10^4$ m⁻¹ corresponding to 3.5μm IR illumination, and $\alpha_B = 3 \times 10^6$ m⁻¹, heat exchange coefficient $g_F = 10^3$ W/(K m²). **(c, d)** Variation of the surface vertical displacement $u_3(x)$ calculated for different light absorption coefficients $\alpha_A = 3.5 \times 10^4$ m⁻¹ and $\alpha_B = (10^4 - 10^6)$ m⁻¹ (different curves in the plot **(c)**) and different heat exchange coefficients $g_F = (10^3 - 10^5)$ W/(K m²) (different curves in the plot **(d)**). Other parameters are given in **Table I**.

The dependence of the doped PMMA surface displacement profile on the distance to the nominal boundary AB (x = 0) has the form of a smoothed step [see **Fig. 3 (c)** and **3 (d)**]. The height and width of the step depend substantially on the difference of the absorption coefficients $\alpha_A - \alpha_B$ [see different curves in **Fig. 3 (c)**]. The step is absent at equal coefficients $\alpha_A = \alpha_B$ (the doping level of both materials is the same in the case), and becomes noticeable (the height is about 10 pm) for a coefficient difference of 3 times [a relatively small difference in the degree of doping in regions A and B]. When the absorption coefficients differ in 50 times (region B is doped and region A is almost pure), the step height saturates (at the level about (100 – 200) pm), while the step width (about 2 μm) is determined by the smallest of the absorption coefficients. The height and profile of the step depends on the value of the heat exchange coefficient $g_F$, and the lowest step with the height of 100 pm corresponds to the highest $g_F=3\times10^5$ W/(K m$^2$), and the highest step with the height of 200 pm corresponds to the smallest $g_F=3\times10^3$ W/(K m$^2$) (see different curves in **Fig. 3 (d)**).

The numerical results shown in **Figs. 3 (c)** and **3 (d)** for thin films are qualitatively and in part quantitatively consistent with approximate analytical calculations for sufficiently thick films, the results of which are shown in **Fig. 2 (a)** and **2 (b)**. This gives us grounds to regard that the conclusions concerning the peculiarities of the thermo-elastic response near the boundary of two layers, made on the basis of the linear theory in the decoupling approximation between the thermal and elastic field, are qualitatively valid in the case of a nonlinear thermo-elastic response of the films of arbitrary thickness, if all the difference in media A and B is in the light absorption coefficients.

Using the continuum media theory of heat conductivity and elasticity we performed analytical calculations and finite element modeling (FEM) of the thermoelastic response across the boundary between two layers with different light absorption, thermal and elastic parameters. Analytical results derived in the decoupling approximation show that the thermoelastic response amplitude is proportional to the intensity of the illuminating light and thermal expansion coefficient. The width of the thermoelastic response transient region is proportional to the sum of inverse light adsorption coefficients and the to effective thermal transfer length. The difference of displacements height across the T-shape boundary ("thermo-elastic step") is proportional to the difference of the adsorption coefficients and inversely proportional to the heat transfer coefficient. The step height becomes thickness-independent for thick films and proportional to $h^2$ in a very thin film. Numerical results of FEM performed for the boundary between pure and dope PMMA on Si substrate approve that analytical results are qualitatively valid for the layers of arbitrary thickness, and are valid semi-quantitatively when the layers differ only by the light adsorption coefficients.

**Supplementary Materials**, which include calculations details

**Acknowledgement:** This manuscript has been authored by UT-Battelle, LLC, under Contract No. DE-AC0500OR22725 with the U.S. Department of Energy. The part of the work (CVK, OO), is performed at the Center for Nanophase Materials Sciences, supported by DOE SUFD.

# Supplementary Materials
## APPENDIX A. Problem statement

**A. Heat sub-problem.** Temperature distributions in the multi-region system is described by a coupled system of linear thermal conductivity equations, at that each layer is characterized by its own equation for the temperature variation $\vartheta_m(x,z,t)$ inside each region:

$$c_A \frac{\partial}{\partial t}\vartheta_A = k_A^T\left(\frac{\partial^2}{\partial x^2}+\frac{\partial^2}{\partial z^2}\right)\vartheta_A + q_A(x,z,t), \qquad \text{(region A at } t\text{=0: } x\leq 0,\ 0\leq z\leq h\text{)} \qquad (1a)$$

$$c_B \frac{\partial}{\partial t}\vartheta_B = k_B^T\left(\frac{\partial^2}{\partial x^2}+\frac{\partial^2}{\partial z^2}\right)\vartheta_B + q_B(x,z,t), \qquad \text{(region A at } t\text{=0: } x\geq 0,\ 0\leq z\leq h\text{)} \qquad (1b)$$

$$c_C \frac{\partial}{\partial t}\vartheta_C = k_C^T\left(\frac{\partial^2}{\partial x^2}+\frac{\partial^2}{\partial z^2}\right)\vartheta_C + q_C(x,z,t). \qquad \text{(substrate C: } -\infty < x < \infty,\ z > h\text{)} \qquad (1c)$$

Note that the physical boundary $S_{AB}$ between the regions A and B can shift with time as well as the top surface $z=0$ bends due to the thermo-elastic effects, but in order to derive analytical results we will neglect the effect in the thermal problem solution at the first approximation [see **Fig. 1(a)** and **(b)**]. Numerical simulations using e.g. FEM will account for all these changes in a self-consistent way.

The coefficients $\kappa_m = k_m^T/c_m$, where $c_m$ is the heat capacity and $k_m^T$ is the thermal conductivity of the region "$m$", where $m$=A, B or $C$. Assuming that all thermal sources $q_m(x,z,t)$ are proportional to the light intensity distribution, $I_m(x,z,t) = I_0(x,t)\exp(-\alpha_m z)$ given by the Beer law, $q_m(x,z,t) \sim I_m(x,z,t)$. Eventually

$$q_m(x,z,t) = \gamma_m I_0(x,t)\exp(-\alpha_m z) \qquad (z\geq 0) \qquad (2)$$

Being not interested in the transient process, we can regard that $I_0(x,t) = I_0(x)(1+\delta\sin(\omega_0 t))$, where the constant $|\delta| \ll 1$; light absorption coefficients $\alpha_m$ and prefactors $\gamma_m I_0$ are different in the regions A, B and C located at $z\geq 0$.

Since the relation between the heat flux and the temperature variation is given by the conventional expression, $\vec{j}_m = -k_m^T \left.\frac{\partial \vartheta_m}{\partial \vec{n}}\right|_{S_m}$, the thermal boundary conditions to Eqs. (1a-b) on the top surface with initial location $z=0$ have the form of heat balance equations with different heat exchange coefficients $g_m$ between the top surfaces of the regions A and B and environment:

$$j_3^A = -k_A^T \left.\frac{\partial \vartheta_A}{\partial \vec{n}}\right|_{S_A} = -g_A \vartheta_A(x,z,t), \quad j_3^B = -k_B^T \left.\frac{\partial \vartheta_B}{\partial \vec{n}}\right|_{S_B} = -g_B \vartheta_B(x,z,t), \tag{3a}$$

The boundaries initial location are $S_A = \{x \leq 0, \ z=0\}$ and $S_B = \{x \geq 0, \ z=0\}$

The thermal boundary conditions at the physical boundary $S_{AB}$ are the continuity of heat fluxes and the equality of the layers' temperatures,.

$$\left. k_A^T \frac{\partial \vartheta_A}{\partial \vec{n}} - k_B^T \frac{\partial \vartheta_B}{\partial \vec{n}} \right|_{S_{AB}} = 0, \quad \vartheta_A - \vartheta_B \big|_{S_{AB}} = 0 \tag{3b}$$

The boundary initial location is $S_{AB} = \{x=0, \ 0 \leq z \leq h\}$.

The boundary conditions at the substrate ($z=h$) are similar to (3b), namely:

$$\left. k_A^T \frac{\partial \vartheta_A}{\partial z} - k_C^T \frac{\partial \vartheta_C}{\partial z} \right|_{S_{AC}} = 0, \quad \vartheta_A - \vartheta_C \big|_{S_{AC}} = 0 \tag{3c}$$

$$\left. k_B^T \frac{\partial \vartheta_B}{\partial z} - k_C^T \frac{\partial \vartheta_C}{\partial z} \right|_{S_{BC}} = 0, \quad \vartheta_B - \vartheta_C \big|_{S_{BC}} = 0 \tag{3d}$$

The boundary locations are $S_{AC} = \{x \leq x_{AB}, \ z=h\}$ and $S_{BC} = \{x \geq x_{AB}, \ z=h\}$, with initial location $x_{AB} = 0$ at that only position $x_{AB}$ may changes with time, but for a thick film $x_{AB} \approx 0$.

**B. Thermoelastic sub-problem.** Generalized linear Hooke's law relating elastic stresses and strains in a thermo-elastic problem for each region A, B, C has the form

$$\sigma_{pq} = c_{pqij}^m \left( u_{ij} - \beta_{ij}^m \vartheta_m \right) \quad \text{or} \quad u_{ij} = \beta_{ij}^m \vartheta_m + s_{ijkl}^m \sigma_{kl}. \tag{4}$$

where $m$=A, B, C without summation; $u_{ij} = \frac{1}{2}\left(\frac{\partial U_i}{\partial x_j} + \frac{\partial U_j}{\partial x_i}\right)$ is the strain tensor, $c_{pqij}^m$ and $s_{ijkl}^m$ are elastic stiffness and compliances tensors, $\beta_{ij}^m$ is the thermal expansion tensor. Hereinafter $x = x_1$, $y = x_2$ and $z = x_3$.

We note that the typical contact area in SPM experiment is well below micron-scale. The corresponding intrinsic resonance frequencies of material are thus in the GHz range, well above the practically important limits (both in terms of ion dynamic, and SPM-based detection of localized mechanical vibrations). Hence, we consider the general equation of mechanical equilibrium in the quasi-static case. This mechanical equilibrium equation, $\partial \sigma_{ij}/\partial x_j = 0$, leads to the Lame-type equations for mechanical displacement vector $U_i$ in the regions A, B, C:

$$c_{ijkl}^m \left( \frac{\partial^2 U_k}{\partial x_j \partial x_l} - \beta_{kl}^m \frac{\partial \vartheta_m}{\partial x_j} \right) = 0. \tag{5}$$

$$c_{ijkl} \frac{\partial^2 u_k}{\partial x_j \partial x_l} - c_{ijkl} \cdot \beta_{kl} \frac{\partial \delta C}{\partial x_j} = 0$$

Where the gradient of thermal strains causes the bulk forces with components $F_k = c_{ijpk} \beta_{ij}^m \frac{\partial \vartheta_m}{\partial x_p}$.

Elastic boundary conditions to Eqs.(5) are the absence of the normal stresses mechanically free surfaces $S_A$ and $S_B$ of media "A" and "B"

$$\left( \sigma_{pq} n_q \right)\big|_{S_A} \equiv c_{pqij}^A n_q \left( u_{ij}^A - \beta_{ij}^A \vartheta_A \right)\big|_{S_A} = 0, \quad \left( \sigma_{pq} n_q \right)\big|_{S_B} \equiv c_{pqij}^B n_q \left( u_{ij}^B - \beta_{ij}^B \vartheta_B \right)\big|_{S_B} = 0 \tag{6a}$$

At the interface between "A" and "B" conditions of continuity of normal stresses $\left( \sigma_{pq} n_q \right)\big|_{S_{AB}-0} = \left( \sigma_{pq} n_q \right)\big|_{S_{AB}+0}$ and mechanical displacement $U_i\big|_{S_{AB}-0} = U_i\big|_{S_{AB}+0}$ should be satisfied:

$$\left( c_{pqij}^A \left( u_{ij}^A - \beta_{ij}^A \vartheta_A \right) - c_{pqij}^B \left( u_{ij}^B - \beta_{ij}^B \vartheta_B \right) \right) n_q \big|_{S_{AB}} = 0, \quad \left( U_i^A - U_i^B \right)\big|_{S_{AB}} = 0 \tag{6b}$$

At the boundary with matched or rigid substrate elastic displacements should be continuous or zero, respectively:

$$U_i^A\big|_{S_{AC}} = U_i^B\big|_{S_{BC}} \quad \text{(matched substrate)} \quad U_i^A\big|_{S_{AC}} = U_i^B\big|_{S_{BC}} = 0 \quad \text{(rigid substrate)} \tag{6c}$$

The boundary locations are $S_{AC} = \{x \leq x_{AB}, \ z = h\}$ and $S_{BC} = \{x \geq x_{AB}, \ z = h\}$, with initial location $x_{AB} = 0$ at that only position $x_{AB}$ may changes with time, but for a thick film $x_{AB} \approx 0$.

## APPENDIX B. Analytical solution in decoupling approximation
### A. Analytical solution of elastic sub-problem

Analytical solution of elastic sub-problem (5)-(6) is possible in decoupling approximation from the thermal problem and assuming that all elastic properties of the media A and B are the same (i.e. $c_{pqij}^A \approx c_{pqij}^B$), and both materials are placed on elastically matched substrate (i.e. the latter assumption is a very good approximation for the case when the materials A and B are in fact one material doped by different photoactive impurities (including the particular cases of one doped and other pristine materials). In these approximations the elastic boundary between the media A and B virtually disappears, only boundary conditions (6a) and (6c) matters at $x=0$ and $x=h$. We further restrict the analysis to the transversally isotropic thermal expansion tensor $\beta_{ij} = \delta_{ij}\beta_{ii}$ with $\beta_{11} = \beta_{22} \neq \beta_{33}$ ($\delta_{ij}$ is the Kroneker delta symbol).

For the case the general solution of the elastic sub-problem (5) is given by Eqs.(3)-(5) in Ref.[20]. The maximal surface displacement corresponding to the point $z=0$, i.e. surface displacement at the tip-surface junction detected by SPM electronics, is

$$u_3(x_1, x_2, t) = -\frac{1}{2\pi}\iiint_V \left( \beta_{11} \frac{\xi_3\left(2(1+\nu)\left((x_1-\xi_1)^2 + (x_2-\xi_2)^2\right) - (1-2\nu)\xi_3^2\right)}{\left((x_1-\xi_1)^2 + (x_2-\xi_2)^2 + \xi_3^2\right)^{5/2}} + \frac{3\xi_3^3 \beta_{33}}{\left((x_1-\xi_1)^2 + (x_2-\xi_2)^2 + \xi_3^2\right)^{5/2}} \right) \vartheta(\xi_1, \xi_2, \xi_3, t)\, d\xi_1 d\xi_2 d\xi_3 \quad (7a)$$

where $\nu$ is the Poisson coefficient.

After Fourier transformation and using Percival theorem Eq.(7) becomes

$$u_3(x_1, x_2, t) = -\int_{-\infty}^{\infty} dk_2 \int_{-\infty}^{\infty} dk_1 \int_0^h d\xi_3 \left( \begin{array}{c} \exp(-ik_1 x_1 - ik_2 x_2 - k\xi_3) \times \vartheta(k_1, k_2, \xi_3, t) \\ \times\left( \frac{\beta_{33}}{2\pi}(1 + k\xi_3) + \frac{\beta_{11}}{2\pi}(1 + 2\nu - k\xi_3) \right) \end{array} \right). \quad (7b)$$

Here $k = \sqrt{k_1^2 + k_2^2}$, $\vartheta(k_1, k_2, \xi_3, t)$ is the 2D Fourier image of the temperature field $\vartheta(x_1, x_2, \xi_3, t)$ in the film.

For the isotropic thermal expansion tensor, $\beta_{11} = \beta_{22} = \beta_{33} = \beta$, Eq.(7) reduces to:

$$u_3(x_1, x_2, 0, t) = -\frac{(1+\nu)}{\pi}\iiint_V \frac{\xi_3 \beta \vartheta(\xi_1, \xi_2, \xi_3, t)}{\left((x_1-\xi_1)^2 + (x_2-\xi_2)^2 + \xi_3^2\right)^{3/2}} d\xi_1 d\xi_2 d\xi_3$$
$$= -\frac{1+\nu}{\pi}\beta \int_{-\infty}^{\infty} dk_2 \int_{-\infty}^{\infty} dk_1 \int_0^h d\xi_3 \exp(-ik_1 x_1 - ik_2 x_2 - k\xi_3) \vartheta(k_1, k_2, \xi_3, t) \quad (8)$$

Equations (7)-(8) define the surface displacement at location (0,0) induced by the redistribution of temperature defined by $\vartheta(x_1, x_2, x_3, t)$ field. In the next sub-section we consider the case when the temperature field can be found analytically.

**B. Thermo-elastic response of the film with arbitrary thickness**

Appeared that the analytical solution of the thermal sub-problem (1)-(3) is possible only when all thermal properties of the media A and B are the same ($c_A \approx c_B \approx c_F$, $k_A^T \approx k_B^T \approx k_F^T$, $g_A \approx g_B \approx g_F$), except for the light adsorption coefficients ($\alpha_A \neq \alpha_B$) and factors ($\gamma_A \neq \gamma_B$), and semi-infinite substrate that does not contain heat sources (i.e. it is transparent for the light, $\gamma_C = 0$). For the case the thermal boundary condition at the boundary $S_{AB}$ (3c) does not matter, and the T-type geometry of the thermal problem reduces to the simple two-layer problem with inhomogeneous thermal source. Namely

$$c_F \frac{\partial}{\partial t} \vartheta_F = k_F^T \left( \frac{\partial^2}{\partial x^2} + \frac{\partial^2}{\partial z^2} \right) \vartheta_F + q(x,z,t), \qquad \text{(film: } -\infty < x \leq \infty, \ 0 \leq z \leq h) \qquad (9a)$$

$$c_C \frac{\partial}{\partial t} \vartheta_C = k_C^T \left( \frac{\partial^2}{\partial x^2} + \frac{\partial^2}{\partial z^2} \right) \vartheta_C. \qquad \text{(substrate C: } -\infty < x < \infty, \ z > h) \qquad (9b)$$

Here the complex source is

$$q(x,z,t>0) = I_0(t)\exp(-\lambda|x|) \begin{cases} \gamma_A \exp(-\alpha_A z), & x \leq 0, \\ \gamma_B \exp(-\alpha_B z), & x > 0. \end{cases} \qquad (9c)$$

Where $I_0(t) = I_0(1 + \delta \exp(i\omega_0 t))$ e.g. The doping/pump x-profile is very smooth, $\lambda \ll \alpha_A$. Boundary conditions

$$g_F \vartheta_F(x,z) - k_F^T \left. \frac{\partial \vartheta_F}{\partial z} \right|_{z=0} = 0, \qquad (10a)$$

$$\vartheta_F - \vartheta_C \big|_{z=h} = 0, \qquad k_F^T \frac{\partial \vartheta_F}{\partial z} - k_C^T \left. \frac{\partial \vartheta_C}{\partial z} \right|_{z=h} = 0, \qquad \vartheta_C \big|_{z \to \infty} = 0. \qquad (10b)$$

Since the Fourier transformation over x coordinate of the heat source is

$$\tilde{q}(k,z,\omega) = \tilde{I}_0(\omega) \left( \frac{\gamma_A \exp(-\alpha_A z)}{\sqrt{2\pi}(\lambda + ik)} + \frac{\gamma_B \exp(-\alpha_B z)}{\sqrt{2\pi}(\lambda - ik)} \right). \qquad (11)$$

Where $\tilde{I}_0(\omega)$ is the frequency spectrum of e.g. $I_0(1 + \delta \exp(i\omega_0 t))$. The frequency spectrum of the stationary solution of the boundary problem (9)-(10) can be found in the form:

$$\vartheta_F(x,z,\omega) = \int_{-\infty}^{\infty} dk \exp(ikx) \left[ A(k,\omega)\exp(-kz) + B(k,\omega)\exp(kz) + \tilde{\vartheta}_P(k,z,\omega) \right] \qquad (12a)$$

The partial solution $\tilde{\vartheta}_P(\omega, k, z)$ of inhomogeneous Eq.(9a) t satisfies the equation

$$\left(-i\omega\kappa_F + \frac{\partial^2}{\partial z^2} - k^2\right)\tilde{\vartheta}_P = \frac{-\tilde{s}(k,z,\omega)}{k_F^T}, \text{ where } \kappa_F = \frac{c_F}{k_F^T}. \text{ It has the form:}$$

$$\tilde{\vartheta}_P(k,z,\omega) = -\frac{\tilde{I}_0(\omega)}{\sqrt{2\pi}k_F^T}\left(\frac{\gamma_A \exp(-\alpha_A z)}{(\lambda+ik)(\alpha_A^2 - k^2 - i\omega\kappa_F)} + \frac{\gamma_B \exp(-\alpha_B z)}{(\lambda-ik)(\alpha_B^2 - k^2 - i\omega\kappa_F)}\right). \quad (12b)$$

The thermal field of substrate that vanishes at $z \to \infty$ has the form

$$\vartheta_C(x,z,\omega) = \int_{-\infty}^{\infty} dk\, C(k,\omega)\exp(ikx - |k|(z-h)). \quad (12c)$$

The spectral functions $A(k,\omega)$, $B(k,\omega)$ and $C(k,\omega)$ can be found from the boundary conditions (10), which lead to the system of linear equations:

$$A(k,\omega)\left(1 - \frac{k_F^T}{g_F}k\right) + B(k,\omega)\left(1 + \frac{k_F^T}{g_F}k\right) = -\tilde{\vartheta}_P(k,0,\omega) + \frac{k_F^T}{g_F}\frac{\partial\tilde{\vartheta}_P(k,0,\omega)}{\partial z}, \quad (13a)$$

$$A(k,\omega)\exp(-kh) + B(k,\omega)\exp(kh) - C(k,\omega) = -\tilde{\vartheta}_P(k,h,\omega), \quad (13b)$$

$$-kA(k,\omega)\exp(-kh) + kB(k,\omega)\exp(kh) + \frac{k_C^T}{k_F^T}|k|C(k,\omega) = -\frac{\partial\tilde{\vartheta}_P(k,h,\omega)}{\partial z}, \quad (13c)$$

The solution of the system (13) has a rather cumbersome form

$$A(k,\omega) = \left(\frac{\exp(kh)(k_F^T k + k_C^T|k|)\left(k_F^T \frac{\partial\tilde{\vartheta}_P(k,0,\omega)}{\partial z} - g_F\tilde{\vartheta}_P(k,0,\omega)\right)}{\exp(-kh)(g_F + k_F^T k)(k_F^T k - k_C^T|k|) + \exp(kh)(g_F - k_F^T k)(k_F^T k + k_C^T|k|)}\right.$$
$$\left.+ \frac{(g_F + k_F^T k)\left(k_F^T \frac{\partial\tilde{\vartheta}_P(k,h,\omega)}{\partial z} + k_C^T|k|\tilde{\vartheta}_P(k,h,\omega)\right)}{\exp(-kh)(g_F + k_F^T k)(k_F^T k - k_C^T|k|) + \exp(kh)(g_F - k_F^T k)(k_F^T k + k_C^T|k|)}\right), \quad (14a)$$

$$B(k,\omega) = \left(\frac{\exp(-kh)(k_F^T k - k_C^T|k|)\left(k_F^T \frac{\partial\tilde{\vartheta}_P(k,0,\omega)}{\partial z} - g_F\tilde{\vartheta}_P(k,0,\omega)\right)}{\exp(-kh)(g_F + k_F^T k)(k_F^T k - k_C^T|k|) + \exp(kh)(g_F - k_F^T k)(k_F^T k + k_C^T|k|)}\right.$$
$$\left.- \frac{(g_F - k_F^T k)\left(k_F^T \frac{\partial\tilde{\vartheta}_P(k,h,\omega)}{\partial z} + k_C^T|k|\tilde{\vartheta}_P(k,h,\omega)\right)}{\exp(-kh)(g_F + k_F^T k)(k_F^T k - k_C^T|k|) + \exp(kh)(g_F - k_F^T k)(k_F^T k + k_C^T|k|)}\right), \quad (14b)$$

$$C(k,\omega) = A(k,\omega)\exp(-kh) + B(k,\omega)\exp(kh) + \tilde{\vartheta}_P(k,h,\omega). \quad (14c)$$

Finally the Fourier images of the temperature field

$$\tilde{\vartheta}(k,z,\omega) = \begin{cases} A(k,\omega)\exp(-kz) + B(k,\omega)\exp(kz) + \tilde{\vartheta}_P(k,z,\omega), & 0 < z < h, \\ C(k,\omega)\exp(-|k|(z-h)), & z > h. \end{cases} \quad (15)$$

should be substituted in Eq.(7b) to receive the answer of the thermo-elastic problem in the decoupling approximation. Reminding that the thermal solution is y-independent and thus its Fourier image contains $\delta(k_2)$, the substitution leads to

$$u_3(x,\omega) = -\int_{-\infty}^{\infty} dk \int_0^h dz \exp(-ikx - |k|z)\left(\frac{\beta_{33}}{2\pi}(1+|k|z) + \frac{\beta_{11}}{2\pi}(1+2\nu - |k|z)\right)\tilde{\vartheta}(k,z,\omega). \quad (16)$$

Integration over z in Eq.(16) results into the spectral density of effective thermo-elastic response.

### C. Exact solution of the semi-infinite media thermo-elastic response

While the answer (14)-(16) is an analytical expression, it is too cumbersome for further analyses. A possible (but very oversimplified) case is to consider a semi-infinite film ($h \to \infty$) without any substrate for the thermal problem solution. For the case the Fourier image of the temperature field becomes essentially simpler, namely:

$$\tilde{\vartheta}(k,z,\omega) = A(k,\omega)\exp(-kz) + \tilde{\vartheta}_P(k,z,\omega), \quad (17a)$$

$$A(k,\omega) = \frac{1}{g_F + k_F^T|k|}\left(k_F^T \frac{\partial \tilde{\vartheta}_P(k,0,\omega)}{\partial z} - g_F \tilde{\vartheta}_P(k,0,\omega)\right), \quad (17b)$$

Since $\tilde{\vartheta}_P(k,z,\omega)$ is given by Eq.(12b), and its explicit form is

$$\tilde{\vartheta}_P(k,z,\omega) = -\frac{\tilde{I}_0(\omega)}{\sqrt{2\pi}k_F^T}\left(\frac{\gamma_A \exp(-\alpha_A z)}{(\lambda+ik)(\alpha_A^2 - k^2 - i\omega\kappa_F)} + \frac{\gamma_B \exp(-\alpha_B z)}{(\lambda-ik)(\alpha_B^2 - k^2 - i\omega\kappa_F)}\right), \quad (18)$$

the evident form of Eq.(17a) becomes:

$$\tilde{\vartheta}(k,z,\omega) = -\frac{\tilde{I}_0(\omega)}{\sqrt{2\pi}k_F^T}\left(\begin{array}{l}\frac{\gamma_A}{(\lambda+ik)(\alpha_A^2 - k^2 - i\omega\kappa_F)}\left(\exp(-\alpha_A z) - \frac{g_F + k_F^T\alpha_A}{g_F + k_F^T|k|}\exp(-|k|z)\right) + \\ \frac{\gamma_B}{(\lambda-ik)(\alpha_B^2 - k^2 - i\omega\kappa_F)}\left(\exp(-\alpha_B z) - \frac{g_F + k_F^T\alpha_B}{g_F + k_F^T|k|}\exp(-|k|z)\right)\end{array}\right). \quad (19)$$

Substitution of Eq.(19) in Eq.(16), and isotropic approximation for the thermal expansion tensor $\beta_{33} = \beta_{11} = \beta$ and integration over z $\int_0^\infty dz \exp(-ikx - |k|z)\vartheta(k,z,\omega)$ leads to the expression

$$\tilde{u}_3(k,h\to\infty,\omega) = \frac{(1+\nu)\beta I_0(\omega)}{\pi\sqrt{2\pi}k_F^T} \left[ \begin{array}{c} \dfrac{\gamma_A\left(g_F+k_F^T(2|k|+\alpha_A)\right)(\alpha_A^2-k^2)}{2|k|(g_F+k_F^T|k|)(\lambda+ik)(\alpha_A+|k|)^2(\alpha_A^2-k^2-i\omega\kappa_F)} \\ + \dfrac{\gamma_B\left(g_F+k_F^T(2|k|+\alpha_B)\right)(\alpha_B^2-k^2)}{2|k|(g_F+k_F^T|k|)(\lambda-ik)(\alpha_B+|k|)^2(\alpha_B^2-k^2-i\omega\kappa_F)} \end{array} \right] \quad (20)$$

When deriving Eq.(20) we used the identity $2|k|^2 - \alpha_A^2 - \alpha_A|k| = (2|k|+\alpha_A)(|k|-\alpha_A)$.

Note that the analytical expression (20) derived for the semi-infinite film ($h\to\infty$) cannot be directly analyzed in x-domain and compared with numerical modeling for $\omega=0$ (i.e. for the static component), because in the case the poles $\sim 1/|k|$ leads to infinite increase of the displacement field in x-domain at $|x|\gg\alpha^{-1}$. In other words the inverse Fourier transformation of Eq.(20) does not exist in a usual sense at $\omega=0$, however Eq.(20) allows one to estimate the "excess" displacement created by the boundary AB. In the next subsection we analyze approximate expressions for the film s of finite thickness. The original exists at $\omega=\omega_0\neq 0$.

### D. Approximate solution of the film thermo-elastic response. Static responce

If the film thickness $h$ is finite and $\beta_{33}=\beta_{11}=\beta$ the approximate expression for displacement (16) in k-domain can be approximated as:

$$\tilde{u}_3(k,h,\omega) = \frac{(1+\nu)\beta\tilde{I}_0(\omega)}{\pi\sqrt{2\pi}k_F^T} \left( \begin{array}{c} \dfrac{\gamma_A}{(\lambda+ik)}\left(\dfrac{1-\exp[-(\alpha_A+|k|)h]}{(\alpha_A+|k|)(\alpha_A^2-k^2-i\omega\kappa_F)} - \dfrac{g_F+k_F^T\alpha_A}{g_F+k_F^T|k|}\dfrac{1-\exp(-2|k|h)}{2|k|(\alpha_A^2-k^2-i\omega\kappa_F)}\right) + \\ \dfrac{\gamma_B}{(\lambda-ik)}\left(\dfrac{1-\exp[-(\alpha_B+|k|)h]}{(\alpha_B+|k|)(\alpha_B^2-k^2-i\omega\kappa_F)} - \dfrac{g_F+k_F^T\alpha_B}{g_F+k_F^T|k|}\dfrac{1-\exp(-2|k|h)}{2|k|(\alpha_B^2-k^2-i\omega\kappa_F)}\right) \end{array} \right)$$

(21a)

When deriving Eq.(21a) we assumed that the thermal field is not very different from the semi-infinite case [see **Appendix A**].

Being further interested in the case $\omega=0$ for which the seeming pole exists in Eq.(21a), we simplify the expression as

$$\tilde{u}_3(k,h) \approx \frac{(1+\nu)\beta I_0}{\pi\sqrt{2\pi}k_F^T}\frac{1}{(g_F/k_F^T)+|k|}\left(\frac{\gamma_A(1-\exp(-2\alpha_A h))}{2\alpha_A(\lambda+ik)(\alpha_A+|k|)} + \frac{\gamma_B(1-\exp(-2\alpha_B h))}{2\alpha_B(\lambda-ik)(\alpha_B+|k|)}\right) \quad (21b)$$

The expression (21b) can be analyzed and simplified in the case of zero flux on the top surface (i.e. assuming that $g_F \to 0$, corresponding to the absence of heat contact, zero heat flux at z=0) and in the opposite case of zero temperature (i.e. at $k_F^T \to 0$). As one can see the surface displacement (21) either diverges for $g_F \to 0$ due to the pole $\sim 1/|k|$ or becomes very small at $k_F^T \to 0$. Hence the realistic situation when both of the parameters $g_F$ and $k_F^T$ are nonzero should be considered in Eq.(21b). For the case inverse Fourier transformation of Eq.(21b) gives the displacement x-profile in the form:

$$u_3(x,h) = \frac{(1+\nu)\beta I_0}{\pi\sqrt{2\pi}k_F^T}\left[u_f\left(x,h,\gamma_A,\alpha_A,\frac{g_F}{k_F^T},\lambda\right) - u_f\left(x,h,\gamma_B,\alpha_B,\frac{g_F}{k_F^T},-\lambda\right)\right] \quad (22)$$

Where the functions are introduced:

$$u_f\left(x,h,\gamma,\alpha,\frac{g_F}{k_F^T},\lambda\right) = \frac{\gamma}{2\alpha}[1-\exp(-2\alpha h)]\frac{f(x,\alpha,\lambda) - f(x,(g_F/k_F^T),\lambda)}{\alpha - (g_F/k_F^T)} \quad (23a)$$

$$f(x,\alpha,\lambda) = \frac{1}{\sqrt{2\pi}(\alpha^2+\lambda^2)}\begin{bmatrix} 2\mathrm{Ci}(\alpha|x|)(\lambda\cos(\alpha x) - \alpha\sin(\alpha x)) \\ -(\alpha\cos(\alpha x) + \lambda\sin(\alpha x))(\pi\mathrm{sign}(x) - 2\mathrm{Si}(\alpha x)) \\ +\exp(\lambda x)(\pi\alpha(\mathrm{sign}(x)-1) - \lambda[2i\mathrm{Si}(i\lambda x) + (\mathrm{Ci}(-i\lambda|x|) + \mathrm{Ci}(i\lambda|x|))]) \end{bmatrix}$$

(23b)

The cosine integral function Ci(z) is defined as $\mathrm{Ci}(y) = \int_y^\infty dt\frac{\cos(t)}{t}$, the sine integral function Si(z) is defined as $\mathrm{Si}(y) = \int_0^y dt\frac{\sin(t)}{t}$. Despite the complexity, the function $f(x,\alpha,\lambda)$ is real and finite and the approximate expression is valid with high accuracy (less than 1%) at small parameter $\lambda/\alpha_A < 10^{-2}$:

$$f(x,\alpha,\lambda) \approx \frac{1}{\sqrt{2\pi}\alpha}[\pi(\mathrm{sign}(x)-1)\exp(\lambda x) - 2\sin(\alpha x)\mathrm{Ci}(\alpha|x|) - \cos(\alpha x)(\pi\mathrm{sign}(x) - 2\mathrm{Si}(\alpha x))]$$

(24)

Using accurate Pade-type approximations for $\mathrm{Si}(y)$ and $\mathrm{Ci}(|y|)$ are

$$\mathrm{Si}(y) \approx \frac{\pi}{2}\tanh(y) - \frac{y\cos(y)}{y^2+(2/(\pi-2))}, \quad \mathrm{Ci}(|y|) \approx \left(\frac{\gamma + \ln(|y|) - y^2/4 - 1}{1+0.3|y|^3} + 1\right)\frac{\sin(y)}{y}, \quad (25)$$

one can express Eq.(24) via elementary functions, and then lead to the conclusion that the main terms are

$$f(x,\alpha,\lambda) \sim \frac{1}{\sqrt{2\pi}\alpha}\left[\pi(\text{sign}(x)-1)\exp(\lambda x) - \pi\cos(\alpha x)(\text{sign}(x) - \tanh(\alpha x)) - \frac{2\alpha x}{(\alpha x)^2 + (2/(\pi-2))}\right]$$

(26)